\documentstyle[fleqn,11pt]{book}
\begin{document}
\parindent 1.4cm
\large
\begin{center}
{\Large \bf THE ERMAKOV-LEWIS INVARIANTS OF THE GROSS-PITAEVSKII EQUATION}
\end{center}
\begin{center}
{{\bf Jos\'e Maria Filardo Bassalo}}
\end{center}
\begin{center}
{Funda\c{c}\~ao Minerva}
\end{center}
\begin{center}
{{\bf Paulo de Tarso Santos
Alencar}}
\end{center}
\begin{center}
{Professor Aposentado da UFPA}
\end{center}
\begin{center}
{{\bf Daniel Gemaque da Silva}}
\end{center}
\begin{center}
{Professor de Ensino M\'edio\ -\ Amap\'a}
\end{center}
\begin{center}
{{\bf Antonio Boulhosa Nassar}}
\end{center}
\begin{center}
{Extension Program-Department of Sciences, University of California,\
Los Angeles, California 90024}
\end{center}
\begin{center}
{{\bf M. Cattani}}
\end{center}
\begin{center}
{Instituto de F\'{\i}sica da USP,\ 05389-970,\ S\~ao Paulo,\ SP}
\end{center}
\par
ABSTRACT:\ In this work we study the Ermakov-Lewis invariants of
the non-linear Gross-Pitaeviskii equation. \vspace{0.2cm}
\vspace{0.2cm}
\par
1.\ {\bf Introduction}
\par
Many years ago, in 1967[1], H. R. Lewis has shown that there is a
conserved quantity, that will be indicated by I, associated with the
time dependent harmonic oscillator (TDHO) with frequency ${\omega}(t)$,
given by:
\begin{center}
{$I\ =\ {\frac {1}{2}}[({\dot {q}}\ {\alpha}\ -\ {\dot {{\alpha}}}\
q)^{2}\ +\ ({\frac {q}{{\alpha}}})^{2}]$\ ,\ \ \ \ \ (1.1)}
\end{center}
where $q$ and ${\alpha}$ obey, respectively the equations:
\begin{center}
{${\ddot {q}}\ +\ {\omega}^{2}(t)\ q\ =\ 0\ ,\ \ \ \ \ {\ddot
{{\alpha}}}\ +\ {\omega}^{2}(t)\ {\alpha}\ =\ {\frac
{1}{{\alpha}^{3}}}$\ .\ \ \ \ \ (1.2,3)}
\end{center}
\par
On the other hand, as the above expressions have also been obtained by
V. P. Ermakov [2] in 1880, the invariants determination of time
dependent physical systems is also known as the {\bf Ermakov-Lewis
problem}. So, considerable efforts have been devoted to solve this
problem and its generalizations, in the last thirty years, and in many
works have been published on these subjects [3,4].
\par
In the present work we investigate the existence of these invariants
for the one-dimensional non-linear Gross-Pitaevskii equation with the
potencial $V(x,\ t)$ given by:
\begin{center}
{$V(x,\ t)\ =\ {\frac {1}{2}}\ m\ {\omega}^{2}(t)\ x^{2}$\ ,\ \ \ \ \ (1.4)}
\end{center}
which is the time dependent harmonic oscillator potential.
\par
2.\ {\bf Gross-Pitaeviskii Equation}
\par
Em 1961[5,6], E. P. Gross and, independently, L. P. Pitaevskii proposed a
non-linear Schr\"{o}dinger equation to represent time dependent
physical systems, given by:
\begin{center}
{i\ ${\hbar}\ {\frac {{\partial}{\psi}(x,\ t)}{{\partial}t}}\ =\ -\
{\frac {{\hbar}^{2}}{2\ m}}\ {\frac {{\partial}^{2}\ {\psi}(x,\
t)}{{\partial}x^{2}}}\ +\ {\frac {1}{2}}\ m\ {\omega}^{2}(t)\ x^{2}\
{\psi}(x,\ t)\ +\ g{\mid}\ {\psi}(x,\ t)\ {\mid}^{3}$\ ,\ \ \ \ \ (2.1)}
\end{center}
where ${\psi}(x,\ t)$ is a wavefunction and $g$ is a constant.
constante.
\par
Writing the wavefuncition ${\psi}(x,\ t)$ in the polar form, defined by
the Madelung-Bohm transformation[7,8], we get:
\begin{center}
{${\psi}(x,\ t)\ =\ {\phi}(x,\ t)\ e^{i\ S(x,\ t)}$\ ,\ \ \ \ \ (2.2)}
\end{center}
where $S(x\ ,t)$ is the classical action and ${\phi}(x,\ t)$ will be
defined in what follows.
\par
Substituting Eq.(2.2) into Eq.(2.1) and taking the real and imaginary
parts of the resulting equation, we get[9]:
\par
\begin{center}
{${\frac {{\partial}{\rho}}{{\partial}t}}\ +\ {\frac
{{\partial}({\rho}\ v_{qu})}{{\partial}x}}\ =\ 0$\ ,\ \ \ \ \ (2.3)}
\end{center}
\begin{center}
{${\frac {{\partial}v_{qu}}{{\partial}t}}\ +\ v_{qu}\ {\frac
{{\partial}v_{qu}}{{\partial}x}}\ +\ {\omega}^{2}(t)\ x\ =\ -\
{\frac {1}{m}}\ {\frac {{\partial}}{{\partial}x}}\ (V_{qu}\ +\ V_{GP})$\
,\ \ \ \ \ (2.4)}
\end{center}
where:
\begin{center}
{${\rho}(x,\ t)\ =\ {\phi}^{2}(x,\ t)$\ ,\ \ \ \ \ (2.5)\ \ \
(quantum mass density)}
\end{center}
\begin{center}
{$v_{qu}(x,\ t)\ =\ {\frac {{\hbar}}{m}}\ {\frac {{\partial}S(x,\
t)}{{\partial}x}}$\ ,\ \ \ \ \ (2.6)\ \ \ \ \ (quantum velocity)}
\end{center}
\begin{center}
{$V_{qu}(x,\ t)\ =\ -\ {\frac {{\hbar}^{2}}{2\ m}}\ {\frac {1}{{\sqrt
{{\rho}}}}}\ {\frac {{\partial}^{2}{\sqrt
{{\rho}}}}{{\partial}x^{2}}}$\ ,\ \ \ \ \ (2.7)\ \ \ \ \ (Bohm quantum
potential)}
\end{center}
and
\begin{center}
{$V_{GP}\ =\ {\frac {g}{m}}\ {\rho}$\ .\ \ \ \ \ (2.8)\ \ \ \
(Gross-Pitaevskii potential)}
\end{center}
\par
In order to integrate Eq.(2.4) let us assume that the expected value of
{\bf quantum force} is equal to zero for all times $t$, that is,
\begin{center}
{$<\ {\frac {{\partial}V_{qu}}{{\partial}x}}\ >\ \ \ {\to}\ \  0\ \ \
{\Leftrightarrow}\ \ \ {\frac {{\partial}V_{qu}}{{\partial}x}}\
{\vert}_{x\ =\ q(t)}\ ,\ \ \ \ \ <x>\ =\ q(t)$\ .\ \ \ \ \ \ (2.9a-c)}
\end{center}
\par
In this way, we can write Eq.(2.4) into two parts:
\begin{center}
{${\frac {{\partial}v_{qu}}{{\partial}t}}\ +\ v_{qu}\ {\frac
{{\partial}v_{qu}}{{\partial}x}}\ +\ {\omega}^{2}\ x\ =\
k(t)\ [x\ -\ q(t)]$\ ,\ \ \ \ \ (2.10)}
\end{center}
\begin{center}
{$-\ {\frac {1}{m}}\ {\frac {{\partial}}{{\partial}x}}\ (V_{qu}\ +\
V_{GP})\ =\ {\frac {{\partial}}{{\partial}x}}\ ({\frac {{\hbar}^{2}}{2\
m^{2}}}\ {\frac {1}{{\sqrt {{\rho}}}}}\ {\frac {{\partial}^{2}{\sqrt
{{\rho}}}}{{\partial}x^{2}}}\ -\ {\frac {g}{m}}\ {\rho})\ =\ k(t)\ [x\
-\ q(t)]$\ .\ \ \ \ \ (2.11)}
\end{center}
\par
Performing the differentiations indicated in Eq.(2.11) we get,
\begin{center}
{${\frac {{\hbar}^{2}}{4\ m^{2}}}\ [{\frac {1}{{\rho}}}\ {\frac
{{\partial}^{3}{\rho}}{{\partial}x^{3}}}\ -\ {\frac {2}{{\rho}^{2}}}\
{\frac {{\partial}{\rho}}{{\partial}x}}\ {\frac
{{\partial}^{2}{\rho}}{{\partial}x^{2}}}\ +\ {\frac {1}{{\rho}^{3}}}\
({\frac {{\partial}{\rho}}{{\partial}x}})^{3}]\ +\ {\frac {g}{m}}\
{\frac {{\partial}{\rho}}{{\partial}x}}\ =\ k(t)\ [x\ -\ q(t)]$\ .\ \ \
\ \ (2.12)}
\end{center}
\par
To integrate Eq.(2.12) it is necessary to known the initial condition
for ${\rho}(x,\ t)$. Let us assume that for $t\ =\ 0$ the physical
system is represented by a normalized Gaussian wave packet, centered at
$q(0)$, that is,
\begin{center}
{${\rho}(x,\ 0)\ =\ [{\pi}\ {\sigma}(0)]^{-\ 1/2}\ e^{-\ {\frac
{[x\ -\ q(0)]^{2}}{{\sigma}(0)}}}\ =\ {\frac {1}{{\sqrt
{A}}}}\ e^{-\ {\frac {B^{2}}{C}}}$\ ,\ \ \ \ \ (2.13)}
\end{center}
\begin{center}
{$A\ =\ {\pi}\ {\sigma}(0)\ ,\ \ \ \ \ B\ =\ x\ -\ q(0)\ ,\ \ \ \ \
C\ =\ {\sigma}(0)$\ .\ \ \ \ \ (2.14-16)}
\end{center}
\par
Since Eq.(2.13) is a particular solution of Eq.(2.12), we must have:
\begin{center}
{${\frac {{\hbar}^{2}}{4\ m^{2}}}\ [{\frac {1}{{\rho}(x,\ 0)}}\ {\frac
{{\partial}^{3}{\rho}(x,\ 0)}{{\partial}x^{3}}}\ -\ {\frac
{2}{{\rho}^{2}(x,\ 0)}}\ {\frac {{\partial}{\rho}(x,\
0)}{{\partial}x}}\ {\frac {{\partial}^{2}{\rho}(x,\
0)}{{\partial}x^{2}}}\ +\ {\frac {1}{{\rho}(x,\ 0)^{3}}}\ ({\frac
{{\partial}{\rho}(x,\ 0)}{{\partial}x}})^{3}]\ +\ {\frac {g}{m}}\ {\frac
{{\partial}{\rho}(x,\ 0)}{{\partial}x}}\ =$}
\end{center}
\begin{center}
{$=\ k(t)\ [x\ -\ q(t)]$\ .\ \ \ \ \ (2.17)}
\end{center}
\par
Performing the differentiation indicated above, Eq.(2.17) becomes:
\begin{center}
{$[{\frac {{\hbar}^{2}}{m^{2}\ {\sigma}^{2}(0)}}\ -\ {\frac {2\
g}{{\sigma}(0)\ m\ {\sqrt {{\pi}\ {\sigma}(0)}}}}]\ [x\ -\ q(0)]\ =\
k(0)\ [x\ - q(0)]\ \ \ {\to}$}
\end{center}
\begin{center}
{$k(0)\ =\ [{\frac {{\hbar}^{2}}{m^{2}\ {\sigma}^{2}(0)}}\ -\ {\frac {2\
g}{{\sigma}(0)\ m\ {\sqrt {{\pi}\ {\sigma}(0)}}}}]$\ .\ \ \ \ \ (2.18)}
\end{center}
\par
Comparing Eq.(2.18) with the Eqs.(2.12) and (2.13), by analogy we get,
\begin{center}
{$k(t)\ =\ [{\frac {{\hbar}^{2}}{m^{2}\ {\sigma}^{2}(t)}}\ -\ {\frac {2\
g}{{\sigma}(t)\ m\ {\sqrt {{\pi}\ {\sigma}(t)}}}}]$\ ,\ \ \ \ \ (2.19)}
\end{center}
\begin{center}
{${\rho}(x,\ t)\ =\ [{\pi}\ {\sigma}(t)]^{-\ 1/2}\ e^{-\ {\frac {[x\ -\
q(t)]^{2}}{{\sigma}(t)}}}$\ ,\ \ \ \ \ \ (2.20)}
\end{center}
where,
\begin{center}
{${\sigma}^{2}(t)\ =\ {\frac {{\hbar}^{2}}{m^{2}}}\ {\frac {1}{k(t)\ +\
{\frac {2\ g}{{\sigma}(t)\ m\ {\sqrt {{\pi}\ {\sigma}(t)}}}}}}$\ .\ \ \ \
\ (2.21)}
\end{center}
\par
Taking into account Eqs.(2.19-21), let us perform the following
differentiations, remembering that $t$ and $x$ are independent variables:
\begin{center}
{${\frac {{\partial}{\rho}}{{\partial}t}}\ =\ -\ {\frac {1}{2}}\ {\frac
{{\dot {{\sigma}}}}{{\sigma}}}\ {\rho}\ +\ {\rho}\ [{\frac {(x\ -\
q)^{2}}{{\sigma}^{2}}}\ {\dot {{\sigma}}}\ +\ {\frac {2\ (x\ -\
q)}{{\sigma}}}\ {\dot {q}}\ ]$\ ,\ \ \ \ \ (2.22)}
\end{center}
\begin{center}
{${\frac {{\partial}{\rho}}{{\partial}x}}\ =\ -\ {\frac {2\ (x\ -\
q)}{{\sigma}}}\ {\rho}$\ ,\ \ \ \ \ (2.23)}
\end{center}
\par
>From Eq.(2.3) and Eqs.(2.22-23), results
\begin{center}
{${\frac {{\partial}v_{qu}}{{\partial}x}}\ -\ {\frac {2\ (x\ -\
q)}{{\sigma}}}\ v_{qu}\ =\ {\frac {{\dot {{\sigma}}}}{2\ {\sigma}}}\ -\
{\frac {{\dot {{\sigma}}}}{{\sigma}^{2}}}\ (x\ -\ q)^{2}\ -\ {\frac {2\
(x\ -\ q)}{{\sigma}}}\ {\dot {q}}$\ .\ \ \ \ \ (2.24)}
\end{center}
\par
Defining
\begin{center}
{$p(x,\ t)\ =\ -\ 2\ {\frac {(x\ -\ q)}{{\sigma}}}$\ ,\ \ \ \ \
(2.25)}
\end{center}
\begin{center}
{$r(x,\ t)\ =\ {\frac {{\dot {{\sigma}}}}{2\ {\sigma}}}\ -\ {\frac
{{\dot {{\sigma}}}}{{\sigma}^{2}}}\ (x\ -\ q)^{2}\ -\ {\frac {2\ (x\ -\
q)}{{\sigma}}}\ {\dot {q}}$\ ,\ \ \ \ \ (2.26)}
\end{center}
Eq.(2.24) becomes,
\begin{center}
{${\frac {{\partial}v_{qu}}{{\partial}x}}\ +\ p(x,\ t)\ v_{qu}\ =\
r(x,\ t)$\ ,\ \ \ \ \ (2.27)}
\end{center}
which can be integrated, giving:
\begin{center}
{$v_{qu}\ =\ {\frac {1}{u}}\ [{\int}\ r\ u\ {\partial}x\ +\ c(t)]$\ ,\
\ \ \ \ (2.28)}
\end{center}
where,
\begin{center}
{$u\ =\ exp\ ({\int}\ p\ {\partial}x)$\ .\ \ \ \ \ (2.29)}
\end{center}
\par
Using Eqs.(2.20,28,29), the function $u$ given by Eq.(2.29) is written
us[9]:
\begin{center}
{$u\ =\ exp\ {\Big {(}}\ {\int}\ [-\ {\frac {2\ (x\ -\ q)}{{\sigma}}}]\
{\partial}x\ {\Big {)}}\ =\ ({\pi}\ {\sigma})^{1/2}\ {\rho}$\ .\ \ \ \
\ (2.30)}
\end{center}
\par
In this way, defining $I\ =\ r\ u\ x$, and using Eqs.(2.26,30) we
obtain:
\begin{center}
{$I\ =\ {\int}\ r\ u\ {\partial}x\ =$}
\end{center}
\begin{center}
{$=\ {\int}\ [{\frac {{\dot {{\sigma}}}}{2\ {\sigma}}}\ -\ {\frac
{{\dot {{\sigma}}}}{{\sigma}^{2}}}\ (x\ -\ q)^{2}\ -\ {\frac {2\ (x\ -\
q)}{{\sigma}}}\ {\dot {q}}]\ ({\pi}\ {\sigma})^{1/2}\ {\rho}\
{\partial}x\ \ \ {\to}$}
\end{center}
\begin{center}
{$I\ =\ I_{1}\ -\ I_{2}$\ ,\ \ \ \ \ (2.31)}
\end{center}
where,
\begin{center}
{$I_{1}\ =\ {\int}\ [{\frac {{\dot {{\sigma}}}}{2\ {\sigma}}}\ -\
{\frac {{\dot {{\sigma}}}}{{\sigma}^{2}}}\ (x\ -\ q)^{2}]\ ({\pi}\
{\sigma})^{1/2}\ {\rho}\ {\partial}x$\ ,\ \ \ \ \ (2.32)}
\end{center}
and
\begin{center}
{$I_{2}\ =\ {\int}\ [{\frac {2\ (x\ -\ q)}{{\sigma}}}\ {\dot {q}}]\
({\pi}\ {\sigma})^{1/2}\ {\rho}\ {\partial}x$\ .\ \ \ \ \ (2.33)}
\end{center}
\par
To integrate Eq.(2.32) it is necessary, first to perform the
differentiation[9] shown bellow, where Eq.(2.23) is used:
\begin{center}
{${\frac {{\partial}}{{\partial}x}}\ [{\frac {{\dot {{\sigma}}}}{2\
{\sigma}}}\ (x-\ q)\ {\rho}]\ =\ [{\frac {{\dot {{\sigma}}}}{2\
{\sigma}}}\ -\ {\frac {{\dot {{\sigma}}}\ (x\ -\
q)^{2}}{{\sigma}^{2}}}]\ {\rho}$\ .\ \ \ \ \ (2.34)}
\end{center}
\par
Inserting Eq.(2.34) into Eq.(2.32), results
\begin{center}
{$I_{1}\ =\ ({\pi}\ {\sigma})^{1/2}\ {\rho}\ ({\frac {{\dot
{{\sigma}}}}{2\ {\sigma}}})\ (x\ -\ q)$\ .\ \ \ \ \ (2.35)}
\end{center}
\par
Similarly, to calculate $I_{2}$, seen in Eq.(2.33), we need to use
Eq.(2.23) obtaining,
\begin{center}
{$I_{2}\ =\ -\ ({\pi}\ {\sigma})^{1/2}\ {\dot {q}}\ {\rho}$\ .\ \ \ \ \
(2.36)}
\end{center}
\par
Substituting Eqs.(2.35-36) into Eq.(2.31) we see that,
teremos:
\begin{center}
{$I\ =\ ({\pi}\ {\sigma})^{1/2}\ {\rho}\ [{\frac {{\dot
{{\sigma}}}}{2\ {\sigma}}}\ (x\ -\ q)\ +\ {\dot {q}}]$\ .\ \ \ \ \
(2.37)}
\end{center}
\par
Remembering that the quantum velocity $v_{qu}$ is defined by Eq.(2.28)
and using Eqs.(2.30,37) we verify that $v_{qu}$ can be written as:
\begin{center}
{$v_{qu}\ =\ {\frac {{\dot {{\sigma}}}}{2\ {\sigma}}}\ (x\ -\ q)\ +\ {\dot
{q}}\ +\ {\frac {c(t)}{({\pi}\ {\sigma})^{1/2}\ {\rho}}}$\ .\ \ \ \ \
(2.38)}
\end{center}
\par
Assuming that the mass density ${\rho}\ {\to}\ 0$ when ${\mid}\ x\ {\mid}\
{\to}\ {\infty}$ we verify that the parameter $c(t)$ must be equal to
zero. Consequently, $v_{qu}$ becomes,
\begin{center}
{$v_{qu}(x,\ t)\ =\ {\frac {{\dot {{\sigma}}}(t)}{2\ {\sigma}(t)}}\ [x\
-\ q(t)]\ +\ {\dot {q}}(t)$\ .\ \ \ \ \ (2.39)}
\end{center}
\par
Using the above Eq.(2.39) we calculate the following differentiations,
remembering that $t$ and $x$ as independent variables:
\begin{center}
{${\frac {{\partial}v_{qu}}{{\partial}t}}\ =\ {\frac
{{\partial}}{{\partial}t}}\ [{\frac {{\dot {{\sigma}}}}{2\ {\sigma}}}\
(x\ -\ q)\ +\ {\dot {q}}]\ =$}
\end{center}
\begin{center}
{$=\ {\frac {{\ddot {{\sigma}}}}{2\ {\sigma}}}\ (x\ -\ q)\ -\ {\frac
{({\dot {{\sigma}}})^{2}}{2\ {\sigma}^{2}}}\ (x\ -\ q)\ -\ {\frac
{{\dot {{\sigma}}}}{2\ {\sigma}}}\ {\dot {q}}\ +\ {\ddot {q}}$\ .\ \ \
\ \ (2.40)}
\end{center}
\begin{center}
{${\frac {{\partial}v_{qu}}{{\partial}x}}\ =\ {\frac
{{\partial}}{{\partial}x}}\ [{\frac {{\dot {{\sigma}}}}{2\ {\sigma}}}\
(x\ -\ q)\ +\ {\dot {q}}]\ =\ {\frac {{\dot {{\sigma}}}}{2\
{\sigma}}}$\ .\ \ \ \ \ (2.41)}
\end{center}
\par
Now, adding the factor ${\omega}^{2}\ q$ to Eqs.(2.39-41), we see that
Eq.(2.10) becomes,
\begin{center}
{${\frac {{\ddot {{\sigma}}}}{2\
{\sigma}}}\ (x\ -\ q)\ -\ {\frac {({\dot {{\sigma}}})^{2}}{2\
{\sigma}^{2}}}\ (x\ -\ q)\ -\ {\frac {{\dot {{\sigma}}}}{2\ {\sigma}}}\
{\dot {q}}\ +\ {\ddot {q}}\ +\ [{\frac {{\dot {{\sigma}}}}{2\
{\sigma}}}\ (x\ -\ q)\ +\ {\dot {q}}]\ {\frac {{\dot {{\sigma}}}}{2\
{\sigma}}}$\ +}
\end{center}
\begin{center}
{$+\ {\omega}^{2}\ x\ +\ {\omega}^{2}\ q\ -\ {\omega}^{2}\ q\
=\ ({\frac {{\hbar}^{2}}{m^{2}\ {\sigma}^{2}}}\ -\ {\frac {2\ g}{m\
{\sqrt {{\pi}\ {\sigma}^{3}}}}})\ (x\ -\ q)\ \ \ {\to}$}
\end{center}
\begin{center}
{$[{\frac {{\ddot {{\sigma}}}}{2\ {\sigma}}}\ -\ {\frac {({\dot
{{\sigma}}})^{2}}{4\ {\sigma}^{2}}}\ +\ {\omega}^{2}\ -\ {\frac
{{\hbar}^{2}}{m^{2}\ {\sigma}^{2}}}\ +\ {\frac {2\ g}{m\ {\sigma}\
{\sqrt {{\pi}\ {\sigma}}}}}]\ (x\ -\ q)\ +\ {\ddot {q}}\ +\
{\omega}^{2}\ q\ =\ 0$\ .\ \ \ \ \ (2.42)}
\end{center}
\par
To satisfy Eq.(2.42), the following conditions must be obeyed:
\begin{center}
{${\frac {{\ddot {{\sigma}}}}{2\ {\sigma}}}\ -\ {\frac {({\dot
{{\sigma}}})^{2}}{4\ {\sigma}^{2}}}\ +\ {\omega}^{2}\ -\ {\frac
{{\hbar}^{2}}{m^{2}\ {\sigma}^{2}}}\ +\ {\frac {2\ g}{m\ {\sigma}\
{\sqrt {{\pi}\ {\sigma}}}}}\ =\ 0$\ ,\ \ \ \ \ (2.43)}
\end{center}
\begin{center}
{${\ddot {q}}\ +\ {\omega}^{2}\ q\ =\ 0$\ .\ \ \ \ \ (2.44)}
\end{center}
\par
Putting
\begin{center}
{${\sigma}\ =\ {\frac {{\hbar}}{m}}\ {\alpha}^{2}$\ ,\ \ \ \
\ (2.45)}
\end{center}
we obtain,
\begin{center}
{${\dot {{\sigma}}}\ =\ {\frac {2\ {\hbar}}{m}}\ {\alpha}\
{\dot {{\alpha}}}\ ,\ \ \ \ \ {\ddot {{\sigma}}}\ =\ {\frac {2\
{\hbar}}{m}}\ [({\dot {{\alpha}}})^{2}\ +\ {\alpha}\ {\ddot
{{\alpha}}}]$\ .\ \ \ \ \ (2.46-47)}
\end{center}
\par
Inserting Eqs.(2.45-47) into Eq.(2.43):
\begin{center}
{${\frac {2\ {\hbar}}{m}}\ {\frac {[({\dot {{\alpha}}})^{2}\
+\ {\alpha}\ {\ddot {{\alpha}}}]}{{\frac {2\ {\hbar}}{m}}\
{\alpha}^{2}}}\ -\ {\frac {{\frac {4\ {\hbar}^{2}}{m^{2}}}\
{\alpha}^{2}\ ({\dot {{\alpha}}})^{2}}{{\frac {4\ {\hbar}^{2}}{m^{2}}}\
{\alpha}^{4}}}\ +\ {\frac {2\ g}{{\sqrt {{\frac {{\pi}}{m}}}}\
{\alpha}^{3}}}\ -\ {\frac {{\hbar}^{2}}{m^{2}\ {\frac
{{\hbar}^{2}}{m^{2}}}\ {\alpha}^{4}}}\ +\ {\omega}^{2}\ =\ 0\ \ \ {\to}$}
\end{center}
\begin{center}
{${\ddot {{\alpha}}}\ +\ {\omega}^{2}\ {\alpha}\ +\ {\frac {2\
g}{{\hbar}\ {\alpha}^{2}\ {\sqrt {{\frac {{\pi}\ {\hbar}}{m}}}}}}\ =\
{\frac {1}{{\alpha}^{3}}}$\ .\ \ \ \ \ (2.48)}
\end{center}
\par
Note that, although the above Eq.(2.48) is formally identical to that
obtained by Ermakov [see Eq.(1.3)], they are different due to the
Planck's constant ${\hbar}$ which appears in Eq.(2.48). This is the
same kind of difference found between the classical wave equation
(d'Alembertian) and the quantum Schr\"{o}dinger wave equation.
\par
Finally, eliminating the factor ${\omega}^{2}$ into Eqs.(2.44) and
(2.48) we get,
\begin{center}
{${\ddot {{\alpha}}}\ -\ {\frac {{\ddot {q}}\ {\alpha}}{q}}\ +\ {\frac {2\
g}{{\hbar}\ {\alpha}^{2}\ {\sqrt {{\frac {{\pi}\ {\hbar}}{m}}}}}}\ =\ {\frac
{1}{{\alpha}^{3}}}\ \ \ {\to}\ \ \ {\ddot {{\alpha}}}\ q\ -\ {\ddot
{q}}\ {\alpha}\ +\ {\frac {2\ g\ q}{{\hbar}\ {\alpha}^{2}\ {\sqrt {{\frac
{{\pi}\ {\hbar}}{m}}}}}}\ =\ {\frac {q}{{\alpha}^{3}}}\ \ \ {\to}$}
\end{center}
\begin{center}
{${\frac {d}{dt}}\ ({\dot {{\alpha}}}\ q\ -\ {\dot {q}}\ {\alpha})\ =\
{\frac {q}{{\alpha}^{3}}}\ -\ {\frac {2\ g\ q}{{\hbar}\ {\alpha}^{2}\
{\sqrt {{\frac {{\pi}\ {\hbar}}{m}}}}}}\ \ \ \ {\to}$}
\end{center}
\begin{center}
{$({\dot {{\alpha}}}\ q\ -\ {\dot {q}}\ {\alpha})\ {\frac {d}{dt}}\
({\dot {{\alpha}}}\ q\ -\ {\dot {q}}\ {\alpha})\ =\ ({\frac
{q}{{\alpha}^{3}}}\ -\ {\frac {2\ g\ q}{{\hbar}\ {\alpha}^{2}\
{\sqrt {{\frac {{\pi}\ {\hbar}}{m}}}}}})\ ({\dot {{\alpha}}}\ q\ -\
{\dot {q}}\ {\alpha})\ \ \ {\to}$}
\end{center}
\begin{center}
{${\frac {d}{dt}}\ [{\frac {1}{2}}\ ({\dot
{{\alpha}}}\ q\ -\ {\dot {q}}\ {\alpha})^{2}]\ =\ -\ {\frac
{q}{{\alpha}}}\ {\frac {d}{dt}}\ ({\frac {q}{{\alpha}}})\ +\ {\frac {2\
g\ q}{{\hbar}\ {\sqrt {{\frac {{\pi}\ {\hbar}}{m}}}}}}\ {\frac {({\dot
{q}}\ {\alpha}\ -\ {\dot {{\alpha}}}\ q)}{{\alpha}^{2}}}\ \ \ {\to}$}
\end{center}
\begin{center}
{${\frac {d}{dt}}\ [{\frac {1}{2}}\ ({\dot {{\alpha}}}\ q\ -\ {\dot
{q}}\ {\alpha})^{2}\ +\ {\frac {1}{2}}\ ({\frac {q}{{\alpha}}})^{2}\ ]\
=\ {\frac {2\ g\ q}{{\hbar}\ {\sqrt {{\frac {{\pi}\ {\hbar}}{m}}}}}}\
{\frac {d}{dt}}\ ({\frac {q}{{\alpha}}})\ \ \ {\to}$}
\end{center}
\begin{center}
{${\frac {dI}{dt}}\ =\ {\frac {2\ g\ q}{{\hbar}\ {\sqrt {{\frac {{\pi}\
{\hbar}}{m}}}}}}\ {\frac {d}{dt}}\ ({\frac {q}{{\alpha}}})$\ ,\ \ \ \ \
(2.49)}
\end{center}
where,
\begin{center}
{$I\ =\ {\frac {1}{2}}\ [({\dot {{\alpha}}}\ q\ -\ {\dot {q}}\
{\alpha})^{2}\ +\ ({\frac {q}{{\alpha}}})^{2}]$\ ,\ \ \ \ \
(2.50)}
\end{center}
which represents the {\bf Ermakov-Lewis-Schr\"{o}dinger invariant} of
the time dependent harmonic oscillator ($TDHO$)[9]. In conclusion, we
have shown that the {\bf Gross-Pitaesvskii equation} {\underline {has
not}} an {\bf Ermakov-Lewis invariant} for the $TDHO$.
\begin{center}
{{\bf REFERENCES}}
\end{center}
\par
1.\ LEWIS, H. R. 1967. {\it Physical Review Letters 18}, 510; 636 (E).
\par
2.\ ERMAKOV, V. P. 1880. {\it Univ. Izv. Kiev 20}, 1.
\par
3.\ NASSAR, A. B. 1986. {\it Journal of Mathematical Physics 27}, 755;
2949, and references therein.
4.\ ------ 1986. {\it Physical Review A33}, 2134, and references
therein.
\par
5.\ GROSS, E. P. 1961. {\it Nuovo Cimento 20}, 1766.
\par
6.\ PITAEVSKII, L. P. 1961. {\it Soviet Physics (JETP) 13}, 451.
\par
7.\ MADELUNG, E. 1926. {\it Zeitschrift f\"{u}r Physik 40}, 322.
\par
8.\ BOHM, D. 1952. {\it Physical Review 85}, 166.
\par
9.\ BASSALO, J. M. F., ALENCAR, P. T. S., CATTANI, M. S. D. e
NASSAR, A. B. 2003. {\bf T\'opicos da Mec\^anica Qu\^antica de de
Broglie-Bohm}, EDUFPA.
\end{document}